\documentclass[aps,prl,twocolumn,floatfix,superscriptaddress,showpacs,nofootinbib]{revtex4-1}
\usepackage{graphicx,amsmath,amssymb,color,epsf,times}
\begin{document}
\newcommand{\dsh}{{d_s^{\rm (hub)}}}
\newcommand{\ds}{{d_s}}

\setcounter{page}{1}
\title[]{First passage time for random walks in heterogeneous networks}
\author{S.~\surname{Hwang}}
\affiliation{Department of Physics and Astronomy, Seoul National University, Seoul 151-747, Korea}
\author{D.-S.~\surname{Lee}}
\email{deoksun.lee@inha.ac.kr}
\affiliation{Department of Natural Medical Sciences and Department of Physics, Inha University,
Incheon 402-751, Korea}
\author{B.~\surname{Kahng}}
\email{bkahng@snu.ac.kr}
\affiliation{Department of Physics and Astronomy, Seoul National University, Seoul 151-747, Korea}
\date[]{Received \today}

\begin{abstract}
The first passage time (FPT) for random walks is a key indicator of how fast information diffuses in a given system. 
Despite the role of FPT as a fundamental feature in transport phenomena, its behavior, particularly in heterogeneous networks, is not yet fully understood. Here, we study, both analytically and numerically, the scaling behavior of the FPT distribution to a given target node, averaged over all starting nodes. We find that random walks arrive quickly at a local hub, and therefore,  the FPT distribution shows a crossover with respect to time from fast decay behavior (induced from the attractive effect to the hub) to slow decay behavior (caused by the exploring of the entire system). Moreover, the mean FPT is independent of the degree of the target node in the case of compact exploration. These theoretical results justify the necessity of using a random jump protocol (empirically used in search engines) and provide guidelines for designing an effective network to make information quickly accessible.
\end{abstract}
\maketitle

In the information age, as data are created in abundance and uploaded on the World Wide Web, 
it is crucial to be able to search and access pages quickly. In this context, one might wonder 
which node, hub, or any other component of the web is more quickly accessible for data mining on the web. 
To answer this question, the problem of the first passage time (FPT) by random walk (RW) in complex networks can be 
posed as a conceptual framework. Many studies on the FPT have been carried out (with particular focus on fractals 
or disordered media~\cite{Redner2001,kahng}), demonstrating its crucial dependence on the spectral dimension of the 
structure~\cite{Condamin2007, Benichou2010, Meyer2011}. Moreover, the FPT problem has also been studied on heterogeneous 
networks to better understand the impact of heterogeneity of degrees on transport phenomena. 
However, such studies remain in the early stages of investigation. 
The mean FPT was obtained in a random graph~\cite{Sood} by an effective medium approximation, and 
in deterministic networks using a recurrence relation~\cite{Agliari2009, Zhang2009}. 
However, a general framework for the scaling behavior of 
the FPT in heterogeneous networks has not been constructed yet. 

Whereas previous studies focused on the FPT problem between two fixed nodes, we consider in this study the mean FPT for arriving at a given 
target node averaged over all possible starting nodes in the system; this is referred to as the global FPT (GFPT) problem.
Consideration of this problem may be useful in finding the mean clicking number for reaching a given target web page from any other web page 
for the first time. In fact, this problem was initially identified in the seminal work of Montroll~\cite{Montroll}, which was followed 
by several studies on disordered systems. Recently, this problem was examined on heterogeneous networks; results reveal that the mean GFPT 
shows a sublinear behavior with respect to the system size $N$ in some limited cases. This implies that average RW time steps to reach a target 
node depends weakly on the system size. Here, we obtain the GFPT distribution and the mean GFPT as a function of the degree of the target node, 
the exponent of the degree distribution, and the spectral dimension for general cases of random heterogeneous networks that exhibit power-laws 
degree distributions. The GFPT's finite-size scaling behavior is also obtained. 
The values of the spectral dimension and the exponent of the degree distribution affect the mean GFPT and the GFPT distribution. We present their lowest-order behaviors analytically for all possible cases.  Moreover, we test our analytical solutions with numerical simulations on diverse complex networks including artificial networks such as (3,5) and (1,2) flower models~\cite{berker06} and the BA model~\cite{ba}, and real-world networks such as the World-Wide Web~\cite{www}, Internet~\cite{as}, the protein interaction networks of {\it H. sapiens}~\cite{humanppi} and {\it S. cerevisiae}~\cite{yeastppi}, and the protein folding network~\cite{rao,rios}.

Suppose RW motion occurs on a scale-free network, composed of $N$ nodes and $L$ links,  and 
in which degrees of each node are heterogeneous, and are distributed following a power law $p_d(k)\sim k^{-\gamma}$. 
For the moment, we consider the case of simple graphs, in which the number of links between two nodes 
can be either zero or one, and degree-degree correlation between two connected nodes being absent.
A RWer at a certain node $i$ jumps to one of its $k_i$ neighbor nodes with probability $1/k_i$ in the next time step. This 
process is repeated at each subsequent time step. We are interested in how quickly the RWer arrives at the target node $m$ 
for the first time. Let $F_{mi}(t)$ be the FPT probability distribution from node $i$ to $m$. 
In the steady state~\cite{noh04}, the probability of finding the RWer at node $i$ is given by $k_i/2L$. Averaged over all starting node $i$, 
the distribution of the GFPT $F_m(t)$ to the target $m$ after $t$ time steps from $i$ is represented by 
\begin{eqnarray}
F_m(t) &\equiv& \sum_{i=1}^{N} \frac{k_i}{2L} F_{mi}(t),
\label{eq:GFPTD}
\end{eqnarray}
where $2L \equiv \sum_{j=1}^{N} k_j$. 

The FPT distribution $F_{mi}(t)$ of a RW satisfies the renewal equation \cite{hughesbook}:
\begin{equation}
P_{mi}(t) = \delta_{mi}\delta_{t0} + \sum_{t'=0}^t F_{mi}(t') P_{mm}(t-t'),
\end{equation}
where $P_{mi}(t)$ is the occupation probability of the RWer at node $m$ at time
$t$ which started from node $i$ at time $t=0$.
Then the generating function $\mathcal{F}_{mi}(z) \equiv \sum_t z^t F_{mi}(t)$ is related to 
the generating function $\mathcal{P}_{mi}(z) \equiv \sum_t z^t P_{mi}(t)$ as
\begin{eqnarray}
\mathcal{F}_{mi}(z) &=\left\{
\begin{array}{cc}
1-1/\mathcal{R}_m(z) &  \ {\rm for} \ \ m=i,\\
\mathcal{P}_{mi}(z)/\mathcal{R}_m(z) &  \ {\rm for}\  \ m\ne i.
\end{array}
\right.
\end{eqnarray}
Here, $\mathcal{R}_m(z)$ is the generating function of the return-to-origin (RTO) probability: $\mathcal{R}_m(z)=\sum_t z^t R_m(t)$, where $R_m(t)=P_{mm}(t)$. Then, the generating function of the GFPT distribution $\mathcal{F}_m(z) \equiv \sum_{t=0}^{\infty} F_m(t) z^t$ can be written in a closed form: 
\begin{eqnarray}
\mathcal{F}_m(z) &=& {k_m\over 2L} \left(1-{1\over \mathcal{R}_{m}(z)}\right) + \sum_{i\ne m}{k_i\over 2L} {\mathcal{P}_{mi}(z)\over \mathcal{R}_m (z)}  \nonumber\\
&=& {k_m z\over 2L (1-z)} \frac{1}{\mathcal{R}_m(z)},
\label{eq:fptrto}
\end{eqnarray}
where we used the relations $k_i \mathcal{P}_{mi}(z) = k_m \mathcal{P}_{im}$~\cite{noh04} and $\sum_{i\ne m} \mathcal{P}_{im}(z) = (1-z)^{-1}-\mathcal{R}_m(z)$.
We note that $\mathcal{F}_m(z)$ depends only on $\mathcal{R}_m(z)$.

The mean GFPT for a given target node $m$, defined as $T_m = \sum_t t F_{m}(t)$, is obtained  as the first derivative of $\mathcal{F}_m(z)$ with respect to $z$ at $z=1$. 
Using the fact that the RTO probability is constant, $R_m(\infty) = k_m/(2L)$ as $t\to \infty$, we observe that the generating function $\mathcal{R}_m(z)$ is 
contributed mainly by $k_m /[2L(1-z)]$ in the limit $z\to 1$. Then, we introduce $\mathcal{R}_m^*(z)\equiv \sum_t z^t [R_m(t)-R_m(\infty)] = \mathcal{R}_m(z) - k_m/[2L(1-z)]$, and Eq.(\ref{eq:fptrto}) is rewritten as $\mathcal{F}_m(z) = k_m z/[k_m + 2L(1-z) \mathcal{R}_m^*(z)]$. Then, the mean GFPT is obtained as \begin{eqnarray}
T_m &=& \left.{\partial \over \partial z} \mathcal{F}_m(z)\right|_{z=1} \approx \frac{2L}{k_m} \mathcal{R}^*_m(1) + 1 \nonumber \\
&= & {2L\over k_m} \sum_{t=0}^{\infty}\left( R_m(t) - R_m(\infty) \right) + 1.
\label{eq:gmfpt}
\end{eqnarray}
In fact, this formula was obtained in \cite{noh04,Tejedor2009}, which corresponds to the inverse of the RW centrality defined in Ref.~\cite{noh04}, thus characterizing the potential-like influence on the RW motion that is caused by the degree heterogeneity. 

The relation between the RTO probability and the mean GFPT in Eq.~(\ref{eq:gmfpt}) enables us to derive the 
specific behavior of $T_m$.  In our previous study  \cite{Hwang2012}, it was shown for random scale-free networks, in which the degree-degree correlation is absent, that $R_m(t)$ exhibited crossover behavior as
\begin{equation}
R_m(t) \sim \left\{
\begin{array}{ccc}
t^{-{d_s^{\rm (hub)}/2}} &~~ {\rm for}& \ 1 \ll t \ll t_c(k_m),\\
k_m t^{-{d_s/2}} & ~~{\rm for }&  \  t_c(k_m) \ll t\ll t_x,\\
{k_m \over 2L} & ~~{\rm for}&  \  t\gg t_x,
\end{array}
\right.
\label{eq:RTOcross}
\end{equation}
where $d_s$ is the spectral dimension of a given network, defined using the density function of the eigenvalues $\lambda$ of the Laplacian matrix as $\rho(\lambda)\sim \lambda^{d_s/2-1}$ in the limit $\lambda\to 0$ \cite{rammal}. $d_s^{\rm (hub)} = d_s \frac{\gamma-2}{\gamma-1}$. $t_c(k_m) \sim k_m^{{2(\gamma-1)}/{d_s}}$ is a crossover time between the two power-law behaviors, and $t_x \sim (2L)^{2/d_s}$ is the crossover time to reach the stationary state. It should be noted that $\dsh < d_s$; in particularly when $\gamma \to 2$, $\dsh$ is almost zero, and thus, the RTO probability remains almost constant in time. 
 In this case,  a RWer can be effectively trapped at the hub when the network is scale-free with the degree exponent $\gamma \to 2$.  
When the target node is the hub, we obtain $t_c(k_h)\sim t_x$, using $k_h\sim N^{1/{(\gamma-1)}}$ obtained from the natural cutoff. Thus, the intermediate time region disappears. For some random network ensembles \cite{mendes,janke}, the maximum degree of hub scales differently as $N^{1/(5-\gamma)}$ when $2<\gamma<3$, which is less than natural cutoff. In this particular case, the intermediate time region does not disappear since $t_c(k_h) \sim N^{2(\gamma-1)/[d_s(5-\gamma)]} < t_x \sim N^{2/d_s}$. 

\begin{figure}
\includegraphics[width=8.0cm]{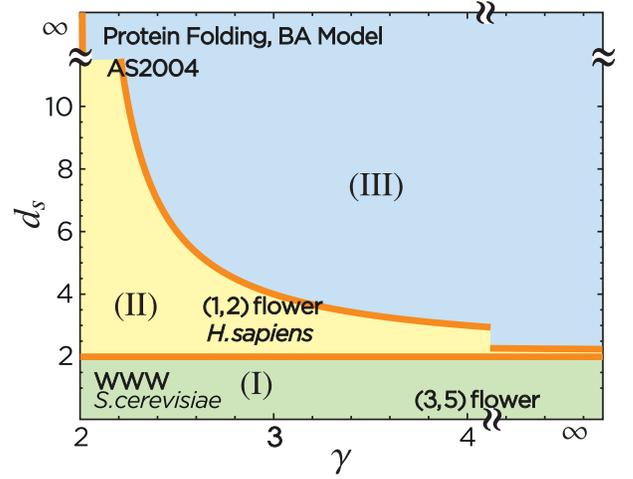}
\caption{(color online). Classification of  networks by the degree exponent $\gamma$ and the spectral dimension $d_s$. 
Three regions are defined as (I) $d_s<2$, (II) $2<d_s<d_c$ where $d_c = 2 (\gamma-1)/(\gamma-2)$, and (III) $d_s>d_c$. 
The networks in different regions show different scaling behaviors in the mean GFPT  as well as the GFPT distribution.}
\label{fig:cases}
\end{figure}

Plugging $R_m(t)$ into Eq.(\ref{eq:gmfpt}), we find that the behavior of the mean GFPT may be classified into three cases, depending on the spectral dimension: 
\begin{eqnarray}
T_m &\approx& \frac{2L}{k_m} \int_{1}^{t_x} [R_m(t) - R_m(\infty)] dt \nonumber\\
&\sim& \left\{
\begin{array}{ccc}
N^{2/d_s} &  \ \rm{(I)} & d_s < 2,\\
N k_m^{-\alpha} &  \ \rm{(II)} & 2 < d_s < d_c ,\\
N k_m^{-1}&  \ \rm{(III)} & d_s>d_c, 
\end{array}
\right.
\label{eq:gmfptScaling}
\end{eqnarray}
where $d_c=2(\gamma-1)/(\gamma-2)$ and $\alpha = \left(1-{2/d_s}\right)(\gamma-1)$. To derive this result, we used $t_x \sim L^{2/d_s}$ and $L \sim N$. 
It is noteworthy that the mean GFPT does not depend on the degree of the target for case (I) of 
compact exploration; however, mean GFPT does depend on the degree of the target for cases (II) and (III) 
in which $d_s > 2$. For example, the spectral dimension of the 
World-Wide Web is $d_s \approx 1.8 < 2$~\cite{Hwang2012}, and thus, the access time to any target node 
is of the same order, irrespective of their degree. That is, the hub would not be a better location for posting information 
than any other node.  
However, for the Internet in the AS system in which $d_s > 2$, the hub can be the most quickly accessible.
For artificial networks such as the Barab\'asi-Albert (BA) model or the static model, $d_s=\infty$,
and thus, the mean GFPT to the hub is the least out of those to all nodes.

We schematically present the regions of the three cases and indicate the region each network occupies   
in the $(\gamma, d_s)$ plane in Fig.~\ref{fig:cases}.  Note that  $\dsh = 2$ locates at the boundary 
between (II) and (III).

\begin{figure}
\includegraphics[width=7.8cm]{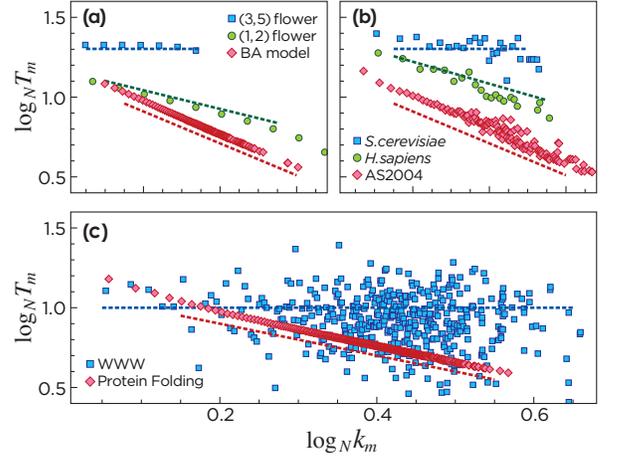}
\caption{(color online). 
Plots of ${\log_N T_m}$ versus ${\log_N k_m}$ are presented. (a) The artificial networks: for the $(3,5)$-flower model with 
$d_s \approx 1.54 < 2$~\cite{rosenfeld,hwang10} in region (I), $T_m$ is independent of $k_m$. However,  for the $(1,2)$-flower model with $d_s \approx 3.17 > 2$~\cite{rosenfeld,hwang10} which is smaller than $d_c\approx 5.44$ and thus in region (II), $T_m$ decays in a power law manner with the exponent $\alpha=(1-2/d_s)(\gamma-1)$, estimated as $\approx 0.23$  (dashed line). For the BA model, $d_s=\infty$~\cite{mendes2} in region (III). $T_m$ decays in a power law manner with an exponent of unity.  The similar plots are drawn for several real-world networks in (b) and (c). 
For a {\it S. cerevisiae} yeast protein interaction network in (b), we obtain $d_s \approx 1.6 < 2$. Together with   
the World Wide Web (c) having $d_s \approx 1.8 < 2$,  the yeast network belongs 
to Region (I) and $T_m$ is independent of $k_m$. 
For a human protein interaction network, we obtain $d_s \approx 2.5 < d_c \approx 3.5$ in region (II), 
$T_m$ decays following a power law with $\alpha \approx 0.43$. Finally, for the AS network (b) and a protein folding network (c),  we obtain $d_s=\infty$ and thus in region (III). $T_m$ decays following a power-law with an exponent of  
unity. All dashed lines are guide lines that were theoretically predicted, and which are close to the numerical 
data.}
\label{fig:gmfpt}
\end{figure}

In Fig.~\ref{fig:gmfpt}, the plot of $\log_N{T_m}$ vs. $\log_N k_m$ is presented for various networks, 
showing different slopes for the networks belonging to the different regions, (I), (II), and (III). 
These slopes are in agreement with the theoretical values indicated by the dashed lines. 
Finally, we note that logarithmic corrections to the mean GFPT can appear at the boundaries between 
the different regions, (I), (II), and (III). 

If the target is the hub, then using the relation $k_h \sim N^{1/(\gamma-1)}$, we obtain that
\begin{eqnarray}
T_h &\sim& \left\{
\begin{array}{cc}
N^{2/d_s} &  \ \rm{(I~~and~~ II)},\\
N^{(\gamma-2)/(\gamma-1)}&  \ \rm{(III)}.
\end{array}
\right.
\label{eq:gmfpt_N}
\end{eqnarray}
Thus, the mean travel time is sublinear for (III), that is,  a RWer can reach the hub without visiting all nodes.
For the case where the degree of hub scales as $k_h \sim N^{1/(5-\gamma)}$ for $2 < \gamma < 3$, we obtain that 
\begin{eqnarray}
T_h &\sim& \left\{
\begin{array}{cc}
N^{2/d_s} &  \ \rm{(I)},\\
N^{[2(3-\gamma)+2(\gamma-1)/d_s]/(5-\gamma)} &  \ \rm{(II)},\\
N^{(4-\gamma)/(5-\gamma)}&  \ \rm{(III)}.
\end{array}
\right.
\label{eq:gmfpt_N1}
\end{eqnarray}

\begin{table*}
\caption{The GFPT distribution for each case (I), (II), and (III). $t_c(k_m) \sim k_m^{{2(\gamma-1)}/{d_s}}$ and $t_x \sim (2N)^{2/d_s}$. $d_s^{\rm (hub)}=d_s\frac{\gamma-2}{\gamma-1}$. $c_1$, $c_2$ and $c_3$ are constants. $\tau_{(\rm I)}\sim N^{d_s/2}$, $\tau_{(\rm II)}\sim Nk_m^{-\alpha}$ and $\tau_{(\rm III)}\sim Nk_m^{-1}$.} 
\scalebox{1.1}{
\begin{tabular*}{0.90\textwidth}{@{\extracolsep{\fill}} c c c c }
\hline
\hline
~~$F_{m}(t)$~~&~~ $1 \ll  t \ll t_c(k_m)$ ~~&~~ $t_c(k_m) \ll t \ll t_x$  ~~&~~ $t\gg t_x$~~ \\
\hline
(I) &
$\frac{k_m}{2N} t^{-(1-d_s^{\rm (hub)}/2)}$ & $\frac{1}{2N} t^{-(1-d_s/2)}$ &$\tau_{\rm (I)}^{-1} \exp( -\frac{ t}{\tau_{\rm (I)}})$ \\
(II) &
$\frac{k_m}{2N} t^{-(1-d_s^{\rm (hub)}/2)}$ &
\multicolumn{2}{c}{$\tau_{\rm (II)}^{-1} \exp( -\frac{t}{\tau_{\rm (II)}})$ }\\
(III)&
\multicolumn{3}{c}{$\tau_{\rm (III)}^{-1} \exp( -\frac{t}{\tau_{\rm (III)}})$ } \\
\hline
\end{tabular*}
}
\label{table:generatingfunctionFPT}
\end{table*}

Next, we solve the long time behavior of the GFPT distribution $F_m(t)$ using Eq.(\ref{eq:fptrto}) and
Eq.(\ref{eq:RTOcross}), which can be determined by $\mathcal{R}_m(z)$ for small $\epsilon=1-z$.
To implement this behavior, we use the approximation $\mathcal{R}_m(z=1-\epsilon) \simeq R_m(\infty)/\epsilon + \int_1^{t_x} (R_m(t)-R_m(\infty)) e^{-\epsilon t} dt$, and determine the leading behavior of $\mathcal{R}_m(z)$ by comparing different time scales $\epsilon^{-1}$, $t_c$, and $t_x$. $\mathcal{R}_m(z)$, which depends on the magnitude of $\epsilon$,  
is determined as follows: 
\begin{eqnarray}
\label{eq:rtoGeneral}
&&\mathcal{R}_m(z=1-\epsilon) - \frac{k_m}{2N \epsilon}   \\ \nonumber
  &\sim&  \left\{
\begin{array}{ccc}
{\rm max}( 1, \epsilon^{{d_s^{\rm (hub)}/2}-1} )~~~~~~~~~~~~~~~~~~~ &  {\rm for} & \ \epsilon_c \ll \epsilon \ll 1,\\
{\rm max}( 1, k_m\epsilon_c^{d_s/2-1} , k_m\epsilon^{{d_s/2}-1}) & {\rm for } & \  \epsilon_x \ll \epsilon \ll \epsilon_c,\\
{\rm max}(1, k_m\epsilon_c^{d_s/2-1}, k_m\epsilon_x^{d_s/2-1}) & {\rm for}  &\   \epsilon \ll \epsilon_x,
\end{array}
\right.
\end{eqnarray}
where
$\epsilon_c =1/t_c(k_m)$ and $\epsilon_x =1/t_x$.

Inserting this result for $\mathcal{R}_m(z=1-\epsilon)$ into Eq.~(\ref{eq:fptrto}), one finds the leading singularity of $\mathcal{F}_m(z=1-\epsilon)$ for small values of $\epsilon$.
Next, applying the Tauberian theorem to $\mathcal{F}_m(z)$ for each case, we obtain $F_m(t)$ as listed in Table I.
We note that the prefactor of $F_m(t)$ in the early-time regime $t\ll t_c(k_m)$ for cases (I) and (II) or $t\ll \tau_{\rm (III)}$ for case (III) is commonly $k_m/(2L)$, suggesting that $F_m(t)$ is proportional to $k_m/N$ using $N\sim L$ for finite $t$. This results because  a RWer far from the target cannot reach it within a finite number of time steps; $F_m(t)$ for finite $t$ is contributed mainly by a RWer who is located close to the target node, which is proportional to the degree of the target node $k_m$.
On the other hand, in the long time regime, for $t\gg \tau_{\rm (I,II,III)}$, the RTO probability converges to the non-zero value $k_m/2L$, which causes the GFPT distribution to decay exponentially in finite networks via the inverse Laplacian 
transformation. The characteristic times $\tau_{\rm (I,II,III)}$ show the same scaling property as the mean GFPT $T_m$ given in Eq.~(\ref{eq:gmfptScaling}).
It is also worth mentioning that for case (I) i.e., $d_s<2$, the exponent in the intermediate time regime $1-d_s/2$ is different from $2-d_s/2$ obtained for the FPT between two given nodes a finite distance apart~\cite{Meroz2011}. This difference implies that a large delay is more likely to occur when receiving information from a source of unknown position. 

\begin{figure}
\includegraphics[width=8.5cm]{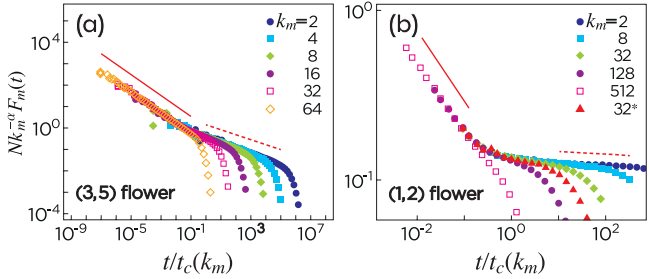}
\caption{
(color online).  Plot of the GFPT distribution in the scaling form for cases (I) (a) and (II) (b). 
The data in (a) are obtained from the (3-5) flower network, in which $d_s \approx 1.54$ and 
$\gamma=4$. Thus, $1-\dsh/2 \approx 0.49$ and $1 - d_s \approx 0.23$ theoretically, 
which is represented by solid and dashed lines, respectively.  
The data in (b) are obtained from the (1-2) flower network with system size $N=29 526$, in which 
$d_s \approx 3.17$ and $\gamma=2.58$. Thus, $1-\dsh \approx 0.42$ theoretically, 
which is represented by solid line. 
To see $N$-dependent behavior of the characteristic time  of $H_{(
\rm II)}(x)$ for $x \gg 1$ in 
Eq.(\ref{eq:scaling2}), we add a dataset of degree $k_m=32$ for a smaller system size $N = 9 843$ denoted with 
the asterisk in the legend. The additional data set are also collapsed well to other data sets in the small regime of 
$x=t/t_c(k_m)$, but it decays at an earlier point $t/t_c(k_m)$ than the corresponding 
point for the larger system $N = 29 526 $. 
}
\label{fig3}
\end{figure}

If the degree $k_m$ of the target node is on the order of one, the crossover time $t_c(k_m)$ is  of order one as well, resulting in the behavior in the intermediate time regime $F_m(t)\sim t^{-(1-d_s/2)}$ for $d_s<2$ and $F_m(t)\sim {\rm const.}$ for $d_s>2$  dominating the whole non-stationary time regime, $t\ll \tau_{\rm (I,II,III)}$. 
In case of (I) and (II), i.e., for $d_s<d_c$,  the different functional behaviors  of the GFPT distributions between the early- and the late-time regimes  can be represented by using a scaling function as
\begin{equation}
F_m^{\rm (I,II)}(t) \sim \frac{k_m^\alpha}{N} H_{\rm (I,II)}\left(\frac{t}{t_c(k_m)}\right),
\label{eq:relation}
\end{equation}
where the scaling function $H_{\rm (I,II)}(x)$ is defined as
\begin{equation}
H_{\rm (I)}(x) = \left\{
\begin{array}{ll}
x^{-1+\dsh/2} & x \ll 1,\\
x^{-1+d_s/2} & x \gg 1,\\
\end{array}
\right.
\label{eq:scaling}
\end{equation}
and 
\begin{equation}
H_{\rm (II)}(x) = \left\{
\begin{array}{ll}
x^{-1+\dsh/2} & x \ll 1,\\
e^{-x/(Nk_m^{1-\gamma})} & x \gg 1.\\
\end{array}
\right.
\label{eq:scaling2}
\end{equation}
The crossover behavior in the scaling form is shown in Fig.~\ref{fig3}. 

Finally, we add two remarks: First, even though our analytic solutions were derived for the networks without the degree-degree correlation, but simulations were carried out for the networks with degree-degree correlation, numerical data are in good agreement with theoretical predictions. This indicates that the degree-degree correlation weakly affects on the mean GFPT and the GFPT distribution. Second, when the network is uncorrelated, maximally 
random networks allowing for multiple edges, maximal degree scales as $k_h \sim N^{1/(\gamma-1)}$ for $\gamma > 2$ \cite{mendes,janke}. In this case,  the mean GFPT to the hub behaves as in Eq. (\ref{eq:gmfpt_N}). 

In summary, we have presented the scaling properties of the mean GFPT and the GFPT distribution analytically for various types of heterogeneous networks. The scaling properties can be classified into three cases depending on the spectral dimension, 
the exponent of the degree distribution, and the degree of a target node. 
Because of the heterogeneity of degrees in the networks,
the mean GFPT displays a sublinear scaling with the system size and the GFPT distribution shows crossover behavior from fast decay behavior to slow decay behavior with respect to time. These properties
can now be used in many applications, for example, search engines on the World-Wide Web, packet transport on the Internet, and protein folding dynamics in biological systems.  

 The authors thank anonymous Referees and Y.W. Kim for helpful comments on the revised manuscript. 
 This work was supported by NRF research grants funded by MEST 
 (Nos.~2010-0015066 (BK) and 2011-0003488 (DSL)). D.-S.L. acknowledges the TJ Park Foundation for support.


\begin{references}
\bibitem{Redner2001} S. Redner, {\sl A Guide to First-Passage Processes} (Cambridge University Press, 2001).
\bibitem{kahng} B. Kahng and S. Redner, J. Phys. A {\bf 22,} 887 (1989).
\bibitem{Condamin2007} S. Condamin, O. Bénichou, V. Tejedor, R. Voituriez, and J. Klafter, Nature {\bf 450}, 77 (2007).
\bibitem{Benichou2010} O. B\'enichou, C. Chevalier, J. Klafter, B. Meyer, and R. Voituriez, Nature Chemistry {\bf 2}, 472  (2010).
\bibitem{Meyer2011} B. Meyer, C. Chevalier, R. Voituriez and O. B\'enichou, Phys. Rev. E {\bf 83}, 051116 (2011).
\bibitem{Sood} V. Sood, S. Redner, and D. ben-Avraham, J. Phys. A {\bf 38}, 109  (2005).
\bibitem{Agliari2009} E. Agliari and R. Burioni, Phys. Rev. E .{\bf 80}, 031125, (2009); 
\bibitem{Zhang2009} Z. Zhang, Y. Qi, S. Zhou, W. Xie, and J. Guan, Phys. Rev. E {\bf 79}, 021127 (2009).
\bibitem{Montroll} E. W. Montroll, J. Math. Phys. {\bf 10,} 753 (1969).
\bibitem{berker06} M. Hinczewski and A. N. Berker, Phys. Rev. E {\bf 73}, 066126 (2006).
\bibitem{ba} A.-L. Barab\'asi and  R. Albert,   Science {\bf 286,} 509 (1999).
\bibitem{www} R. Albert, H. Jeong, and A.-L. Barab\'asi, Nature (London) {\bf 401}, 130 (1999).
\bibitem{as} University of Oregon Route Views Archive Project, http://archive.routeviews.org/.
\bibitem{humanppi} Database of Interacting Proteins, http://dip.doe-mbi.ucla.edu/dip/.
\bibitem{yeastppi} J.-D. Han et al., Nature {\bf 430}, 88 (2004).
\bibitem{rao} F. Rao, and A. Caflisch, J. Mol. Biol {\bf 342,} 299 (2004).
\bibitem{rios} D. Gfeller, P. De Los Rios, A. Caflisch, and F. Rao, Proc. Natl. Acad. Sci. U.S.A {\bf 104,} 1817 (2007).
\bibitem{noh04} J.D. Noh and H. Rieger, Phys. Rev. Lett. {\bf 92}, 118701 (2004).
\bibitem{hughesbook} Huges, B.D. {\sl Random Walks and Random Environments}(Oxford Univ. Press, Clarendon, 1995).
\bibitem{Tejedor2009} V. Tejedor, O. Benichou, and R. Voituriez, Phys. Rev. E {\bf 80}, 065104 (2009).
\bibitem{Hwang2012} S. Hwang, D.-S. Lee, and B. Kahng, Phys. Rev. E {\bf 85}, 046110 (2012).
\bibitem{rammal} R. Rammal and G. Toulouse, J. Physique Lett. {\bf 44}, 13 (1983).
\bibitem{mendes} S.N. Dorogovtsev, J.F.F. Mendes, A.M. Povolotsky, A.N. Samukhin, Phys. Rev. Lett. {\bf 95}, 195701 (2005). 
\bibitem{janke} B. Waclaw, L.Bogacz, W. Janke, Phys. Rev. E {\bf 78}, 061125 (2008).
\bibitem{rosenfeld} H. D. Rosenfeld, S. Havlin, and D. ben-Avraham, New J. Phys. {\bf 9,} 175 (2009). 
\bibitem{hwang10} S. Hwang, C.K. Yun, D.S. Lee, B. Kahng and D. Kim, Phys. Rev. E {\bf 82,} 056110 (2010).
\bibitem{mendes2} A.N. Samukhin, S.N. Dorogovtsev, and J.F.F. Mendes, Phys. Rev. E {\bf 77,} 036115 (2008). 
\bibitem{Meroz2011} Y. Meroz, I. M. Sokolov, and J. Klafter, Phys. Rev. E {\bf 83}, 020104 (2011).
\end{references}
\end{document}